**Transboundary secondary organic aerosol in western Japan indicated by the $\delta^{13}$C of water-soluble organic carbon and the *m/z* 44 signal in organic aerosol mass spectra**

Satoshi Irei,[*,1,†] Akinori Takami,[1] Masahiko Hayashi,[2] Yasuhiro Sadanaga,[3] Keiichiro Hara,[2] Naoki Kaneyasu,[4] Kei Sato,[1] Takemitsu Arakaki,[5] Shiro Hatakeyama,[6] Hiroshi Bandow,[3] Toshihide Hikida,[7] and Akio Shimono[7]

[1]National Institute for Environmental Studies, 16-2 Onogawa, Tsukuba, Ibaraki 305-8506, Japan

[2]Department of Earth System Science, Faculty of Science, Fukuoka University, 8-19-1 Nanakuma, Jonan-ku, Fukuoka 814-0180, Japan

[3]Department of Applied Chemistry, Graduate School of Engineering, Osaka Prefecture University, 1-1 Gakuencho, Naka-ku, Sakai, Osaka 599-8531, Japan

[4]National Institute of Advanced Industrial Science and Technology, 16-1 Onogawa, Tsukuba, Ibaraki 305-8569, Japan

[5]Department of Chemistry, Biology and Marine Science, Faculty of Science, University of the Ryukyus, 1 Senbaru, Nishihara, Okinawa 903-0213, Japan

[6]Agricultural Department, Tokyo University of Agriculture and Technology, 3-5-8 Saiwai-cho, Fuchu, Tokyo 183-8509, Japan

[7]Shoreline Science Research Inc., 3-12-7 Owada-machi, Hachioji, Tokyo 192-0045, Japan

[†]Current address: Department of Chemistry, Biology and Marine Science, Faculty of Science, University of the Ryukyus, 1 Senbaru, Nishihara, Okinawa 903-0213, Japan

*Corresponding author: Satoshi Irei, National Institute for Environmental Studies, 16-2 Onogawa, Tsukuba, Ibaraki 305-8506, Japan (phone: +81-29-850-2314; fax: +81-29-850-2579; e-mail: satoshi.irei@gmail.com)

**ABSTRACT:** The stable carbon isotope ratio ($\delta^{13}C$) of low-volatile water-soluble organic carbon (LV-WSOC) was measured in filter samples of total suspended particulate matter, collected every 24 h in the winter of 2010 at an urban site and two rural sites in western Japan. Concentrations of the major chemical species in fine aerosol (<1.0 μm) were also measured in real time by aerosol mass spectrometers. The oxidation state of organic aerosol was evaluated using $f_{44}$; i.e., the proportion of the signal at $m/z$ 44 ($CO_2^+$ ions from the carboxyl group) to the sum of all $m/z$ signals in the organic mass spectra. A strong correlation between LV-WSOC and $m/z$ 44 concentrations was observed, which suggested that LV-WSOC was likely to be associated with carboxylic acids in fine aerosol. Plots of $\delta^{13}C$ of LV-WSOC versus $f_{44}$ showed random variation at the urban site and systematic trends at the rural sites. The systematic trends qualitatively agreed with a simple binary mixture model of secondary organic aerosol with background LV-WSOC with an $f_{44}$ of ~0.06 and $\delta^{13}C$ of -17‰ or higher.

Comparison with reference values suggested that the source of background LV-WSOC was likely to be primary emissions associated with $C_4$ plants.

**INTRODUCTION**

Organic aerosol (OA) is a major component of atmospheric aerosol.[1] Secondary organic aerosol (SOA), which contains low-volatility oxidation products from the atmospheric oxidation of volatile organic compounds (VOCs), is of significant environmental concern. Our understanding of the oxidation state of OA has advanced through the use of aerosol mass spectrometers (AMSs). An AMS provides $f_{44}$ and $f_{43}$; i.e., the proportions of signal strengths at *m/z* 44 (carboxyl ions) and *m/z* 43 (carbonyl and alkyl ions) in the organic aerosol mass spectra relative to the sum of signal intensities at all *m/z* of the organic mass spectra, respectively.[2,3] High values of $f_{44}$ indicate highly polar and water-soluble OA, as demonstrated by the strong correlation between $f_{44}$ for oxygenated OA and concentrations of water-soluble organic carbon (WSOC) in Tokyo.[4] This correlation alone is still not sufficient to fully understand atmospheric SOA.

Analyses of the stable carbon isotope ratio ($\delta^{13}C$) are potentially useful for characterizing SOA as they indicate the degree of isotope fractionation of precursor VOCs,[5] which allows an estimation of the extent of the oxidation reaction processing.[6] To date, $\delta^{13}C$ measurements have been performed on possible photochemical products, such as formic acid in rainwater[7] and carboxylic acids in aerosols.[8,9] The low $\delta^{13}C$ values have led to their classification as secondary products. It has been reported that even primary organic compounds show a wide range of $\delta^{13}C$ values, including values as low as those reported for secondary products.[10] The features of SOA observed by laboratory studies include not only its highly depleted carbon isotopic composition in $^{13}C$, but also a systematic variation in $\delta^{13}C$ as the precursor oxidation reactions progress.[11,12] The identification of SOA is therefore more certain if a systematic $\delta^{13}C$ variation of a possible product against an indicator throughout the duration of the oxidation reaction is observed. However, to the best of our knowledge, there is no report that such a systematic variation has been observed in the real atmosphere.

Our objective was to conduct $\delta^{13}C$ measurements of possible photochemical products in the atmosphere, and then evaluate variations in $\delta^{13}C$ as a function of an indicator of the progress of the oxidation reaction. We selected low-volatile WSOC (LV-WSOC) as a tracer of photochemical products and the $f_{44}$ obtained by AMS as an indicator of the progress of the oxidation reaction. We conducted field studies in the winter of 2010 at rural and urban sites in western Japan, where the influence of transboundary air pollution from the Asian continent is significant during winter and spring.

**EXPERIMENT**

Field studies were conducted from 6 to 16 December, 2010, at three locations: the Cape Hedo Atmosphere and Aerosol Monitoring Station (26.9°N, 128.3°E) in Okinawa Prefecture, the Fukue atmospheric monitoring station (32.8°N, 128.7°E) in Nagasaki Prefecture, and Fukuoka University (33.6°N, 130.4°E) in Fukuoka Prefecture (Figure 1). Hedo and Fukue are rural sites, whereas Fukuoka is an urban site in a city with a population of ~1.4 million. At all three sites, air quality in winter and spring is often strongly influenced by outflow from the Asian continent.

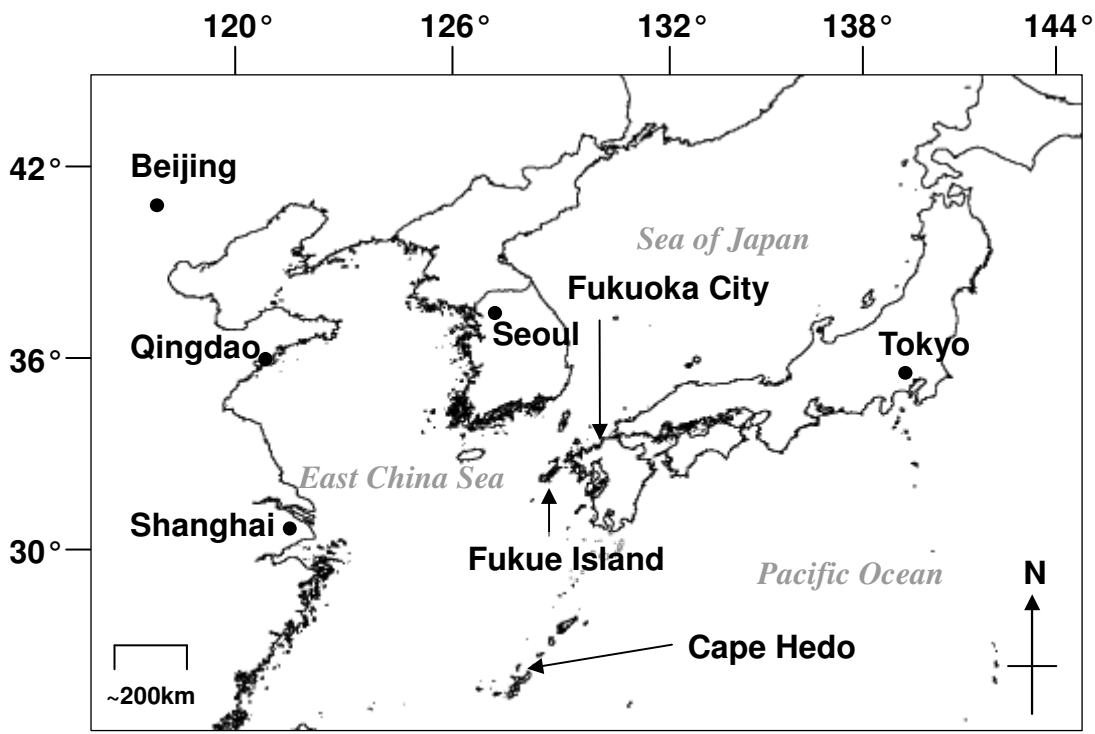

*Figure 1. Map showing the locations of measurement sites.*

At each site, a daily 24-h sample of total suspended particulate (TSP) was collected from noon to noon on an 8 × 10-in. quartz fiber filter (Tissuquartz, Pall Corp., NY, USA) using high-volume air samplers (HV-1000: Sibata Corp., Japan). The samplers were placed on the rooftops of the monitoring stations at Hedo and Fukue (~3-m height) and the rooftop of a building at Fukuoka (~15-m height). Prior to collection, all filters were baked at 773 K for ~12 h to reduce the level of background organic carbon. The sampling flow rate was 1000 L min$^{-1}$, corresponding to a sample volume of 1440 m$^3$ of air for each sample. We collected 11 filter samples and a field blank filter at each site.

Each filter sample was cut into four pieces, and one of these segments was used for the LV-WSOC analysis in a procedure similar to the method of Kirillova et al.,[13] but with a difference in the evaporation stage. WSOC was extracted in a wide-mouth glass jar by ultrasonic agitation for 5 min with 15 mL of reverse-osmosis/ion-exchanged water (RFP542HA: Advantec Inc., Japan). The extract was filtered using a disposable syringe (SS-05SZ: Terumo Corp., Japan) with a 0.45-μm PTFE syringe filter (PURADISC 25TF: Whatman Japan K.K., Japan). This step was repeated two more times with 10 mL of water, and the three extracts were combined. The volume of the extract was reduced to ~0.1 mL in a rotary evaporator (R-205 and B-490: Nihon Büchi K.K., Japan), and then further evaporated in a preweighed conical vial (Mini-vial: GL Sciences Inc., Japan) under a flow of 99.99995% pure nitrogen (Tomoe Shokai, Japan). The volume of the concentrated extract was determined by weighing the vial under the assumption that the extract had a density of 1 g mL$^{-1}$. Then a 0.05–0.1-mL aliquot of the concentrated extract was pipetted (Research Plus, Eppendorf AG,

Germany) into a 0.15-mL tin cup for elemental analysis (Ludi Swiss AG, Switzerland). The extract in the cup was dried under a flow of pure nitrogen. After dryness was confirmed visually, the sample was left under the nitrogen flow for a random interval ranging from 30 to 120 min to prepare for LV-WSOC analysis. A drop of 0.01 M hydrochloric acid (Wako Pure Chemical Industries, Japan) was spiked into the tin cup to remove carbonate, and the extract was dried again. The dried samples prepared in this manner were analyzed by an elemental analyzer (Flash 2000: Thermo Scientific, MA, USA) coupled with an open-split interface (Conflo IV: Thermo Scientific, MA, USA), followed by an isotope ratio mass spectrometer (Delta V Advantage: Thermo Scientific, MA) for determining the carbon mass and $\delta^{13}C$ value, which was defined as follows:

$$\delta^{13}C = \left[ \frac{(\frac{^{13}C}{^{12}C})_{sample}}{(\frac{^{13}C}{^{12}C})_{reference}} - 1 \right],$$

where $(^{13}C/^{12}C)_{sample}$ and $(^{13}C/^{12}C)_{reference}$ are the $^{13}C/^{12}C$ atomic ratios for the sample and the reference (the Vienna Pee Dee Belemnite), respectively,

Analytical tests were conducted with sucrose (IAEA-C6, IAEA, Austria) and oxalic acid (>98% purity, Wako Pure Chemical Industries, Japan) to evaluate recovery yields as well as the accuracy and precision of the isotope measurements. These substances are often the predominant species in WSOC from ambient aerosols.[14,15] The reference value ± standard deviation for the sucrose was −10.8 ± 0.5‰ as recommended by the International Atomic

Energy Agency (IAEA), and for the oxalic acid it was –28.29 ± 0.2‰ ($n$ = 6) as determined in our laboratory by analysis of the pure chemical.

During the study period, we used AMSs to measure the chemical composition of fine aerosol (approximately corresponding to $PM_{1.0}$) at the three sites. Quadrupole AMSs (Aerodyne Research Inc., MA, USA) were used at Fukue and Fukuoka. The details of the instrumentation and the procedure for determining the concentration of chemical species from mass spectra are described elsewhere.[16,17] The time resolution was 10 min with a scan range from $m/z$ 1 to $m/z$ 300. We used an aerosol chemical speciation monitor (ACSM: Aerodyne Research Inc., MA, USA) at Hedo for this analysis. Details of this instrument and its calibration are described elsewhere.[18] The time resolution was 5 min with a scan range from $m/z$ 1 to $m/z$ 150. The difference resulting from the different scan ranges of the AMSs and the ACSM was small because the organic mass concentration was predominantly determined by the mass range less than $m/z$ 100. All instruments had heater temperatures set to 873 K, and all instruments were calibrated with 300–350-nm dried ammonium nitrate particles at the beginning of the study period to determine the ionization efficiency (for the AMSs) or the response factor (for the ACSM) for nitrate. The ionization efficiencies determined for the AMSs were $9.6 \times 10^{-7}$ and $5.5 \times 10^{-7}$ counts molecule$^{-1}$ at Fukue and Fukuoka, respectively, and the response factor for the ACSM was $6.8 \times 10^{-11}$ A m$^3$ µg$^{-1}$ at Hedo. Finally, concentrations of sulfate measured by the AMSs and the ACSM were averaged over 24-h, and those 24-h average concentrations were compared with daily average concentrations of non-sea-salt sulfate (= total sulfate concentration – 0.251 × [Na$^+$]) obtained from an analysis of the TSP filter samples, to optimize the collection efficiencies of the AMSs and the ACSM.

Additionally, mixing ratios for the sum of NO and NO$_2$ (NO$_x$) and total odd nitrogen (NO$_y$) were measured to retrieve another atmospheric oxidation indicator, photochemical age ($t$[OH]). Concentrations of NO$_x$ and NO$_y$ were determined using commercially available NO$_x$ analyzers (Model 42 i-TL: Thermo Scientific, MA, USA). The detailed operation of the analyzers is described elsewhere.[19,20] The atmospheric detection limit for both measurements was about 60 pptv. These measurements were performed only at Fukue.

## RESULTS AND DISCUSSION

**Validation for LV-WSOC Analysis and AMS Measurement.** Measurements of both the recovery test blanks ($n$ = 5), which were handled in the same manner as the recovery test samples, and the field blanks ($n$ = 3), which were handled in the same manner as the field samples, including sample transportation, were of the same magnitude as the random variations. In overall, the average ± standard deviation was 10 ± 5 μgC (range 4 to 18 μgC) for the carbon mass and –23 ± 4‰ (range –16‰ to –27‰) for $\delta^{13}$C. These average blank values were used for blank corrections. Note that the average blank size of 10 μgC was ~14% of the smallest sample size. Taking the detection limit of the analysis as three times the standard deviation of the blank values, the detection limit for carbon mass was approximately 4 µgC, i.e., 3% of the smallest sample size. All LV-WSOC concentrations and $\delta^{13}C$ data for the ambient samples were corrected for the blank value.

Results of the standard spike tests ($n = 5$) showed that the average blank-corrected recovery yield ± standard deviation was 89 ± 11% (range 73% to 102%) for IAEA-C6 and 77 ± 3% (range 74% to 80%) for oxalic acid. No dependency of the recovery yields on evaporation time was observed. Our recovery yields for oxalic acid were better than those reported by Kirillova et al.,[13] which was probably because the evaporation of solvent in our sample preparation was undertaken at atmospheric pressure. The differences between the reference $\delta^{13}C$ values and the blank-corrected $\delta^{13}C$ values were on average +0.4 ± 0.1‰ (range = 0.26‰ to 0.54‰) for IAEA-C6 and +0.4 ± 0.3‰ (range = 0.03‰ to 0.74‰) for oxalic acid. These differences were small, but statistically significant. Although our results may contain small biases of between 23% and 11% for recovery yields and +0.4‰ for $\delta^{13}C$ values, the WSOC concentrations and their $\delta^{13}C$ values reported here were not corrected for bias.

Although correlation plots are not shown here, we found that the 24-h average concentrations of sulfate determined by the AMSs and the ACSM were strongly correlated with the 24-h average concentrations of non-sea-salt sulfate ($r^2 > 0.87$). Based on the slopes of the regression lines, the collection efficiencies were determined to be 1, 0.74, and 1 for Hedo, Fukue, and Fukuoka, respectively. It should be noted that, for the AMS data collected at Fukuoka, the signal at m/z 29 that accounted for only ~7% of total OA was significantly influenced by the signal at m/z 28, due to $N_2$ in the air. For this reason, the signal at m/z 29 was excluded from the data analysis.

**Time Series Variation.** The AMS measurements showed that sulfate, which is common in secondary air pollution from continental China, was the dominant species at the two rural sites, whereas organics were the dominant species at Fukuoka (Figure S-1 in the Supporting Information). These observations at the rural sites are consistent with results from previous studies.[21,22] The time series of sulfate concentrations displayed the same pattern of variation at Fukue and Fukuoka, despite the locations being ~200 km apart, while a different pattern was observed at Hedo. The similarity between Fukue and Fukuoka implies that the same air masses influenced both sites, while the different pattern at Hedo suggests a different origin for the sulfate precursor, $SO_2$. However, it should be noted that there were some periods (e.g., December 7), in which the variations at Fukue and Fukuoka did not correspond. The back trajectories of air masses modeled by HYSPLIT[23] at the three sites showed similar trajectories passing through northeastern China, and sometimes through the Korean peninsula, that were indicative of the possible influence of transboundary pollution (Figure S-2). The variations in the concentration of organics at Fukue and Fukuoka were different. This is likely to be due to irregular contributions of organics from local sources around the Fukuoka site.

Time series plots of LV-WSOC concentrations at Fukue and Fukuoka showed similar variations, whereas the variation at Hedo was clearly different (Figure S-3). This is consistent with the sulfate observations discussed previously. Although a scatter plot is not shown here, we found that the variation in the LV-WSOC concentration was strongly correlated between Fukue and Fukuoka ($r^2$ >0.91), but not between Fukue and Hedo ($r^2$ ~0.02). Furthermore, the concentration of LV-WSOC at Fukue was considerably higher than at Fukuoka, which is consistent with its proximity to the Asian continent as shown by the timing of high-concentration episodes, during which Fukue was affected several hours before Fukuoka. The observations suggest that the majority of LV-WSOC was likely to be of continental origin and that Fukue was more strongly influenced than Fukuoka by outflow from the continent.

Time series plot of hourly average $NO_x$ and $NO_y$ mixing ratios at Fukue are shown in Figure S-4. The $NO_x$ and $NO_y$ mixing ratios displayed a similar variation. The patterns of variation were similar to the variations in the concentrations of chemical species measured by AMS and the LV-WSOC concentrations determined by analysis of the filter samples. The similarity of these variations indicates that their origins are likely to be the same, transboundary pollutants.

**Oxidation of OA measured by AMS.** The *m/z* 44 concentration was strongly correlated with the OA concentration ($r^2 \geq 0.86$) at all sites (Figure S-5). The slopes of the linear regressions were significantly different: $0.23 \pm 0.02$ for Hedo, $0.142 \pm 0.008$ for Fukue, and $0.08 \pm 0.01$ for Fukuoka. The high value for Hedo is probably due to the significant contribution of carboxylic acids, which are major components of the atmospheric aerosol at remote sites[14,15] and are suspected to be of secondary origin. The relatively low value for Fukuoka is probably due to the contribution of primary OA from local emission sources, and the intermediate value for Fukue is likely to be the result of a mixture of relatively fresh OA and SOA, which is consistent with the results of previous field studies conducted at Fukue.[21, 24]

We also conducted positive matrix factorization (PMF) analysis on our AMS data using a PMF evaluation tool.[25] Several issues with PMF analysis should be noted: (1) PMF, like any factor analysis, reliably extracts only significantly contributing factors, and the analysis usually ends with some residuals unaccounted for by these factors; and (2) in factor analysis the choice of the number of factors is strongly dependent on the judgment of the analyst, therefore the user must carefully select the number of factors that best explains the reconstructed results.[25] As a scale of the extent of OA oxidation, Ng et al.,[2] reported that a plot of $f_{44}$ versus $f_{43}$ for oxygenated OA (OOA) factor yielded by PMF analysis fell between two dashed lines ($y = -0.60204x + 0.4154$ and $y = -1.8438x + 0.3319$), which converged at $f_{44}$

= 0.295 and $f_{43}$ = 0.020. The plots of 24-h average $f_{44}$ versus 24-h average $f_{43}$ for fine OA at Hedo, Fukue, and Fukuoka were clustered in the top, middle, and bottom of the triangle, respectively (Figure S-6). It was found that the PMF analyses with one or two factors gave loadings that were in reasonable agreement with the reference mass spectra of low-volatile OOA (LV-OOA) and hydrocarbon-like organic aerosol (HOA) found in the AMS spectral database.[26] The loadings of LV-OOA and HOA yielded at each site are shown in Figures S-7, S-8, and S-9, and their mass fractions in the OA are summarized in Table S-1. Briefly, ~99% of OA at Hedo consisted of a single factor (LV-OOA), which is believed to be of SOA origin. The OA at Fukue consisted mainly of two factors (LV-OOA and HOA), which are recognized as secondary and primary OA, respectively. The OA at Fukuoka also consisted of LV-OOA and HOA, but the fraction of HOA was more significant than that at Fukue. The results of our PMF analysis for the Hedo and Fukue data are consistent with the results of principal-component analyses in previous studies.[24]

We further investigated the composition of deconvolved LV-OOA using an elemental analysis technique based on signal intensities at selected m/z for specific fragment ions.[26] This analysis gives the relative elemental composition of C, H, and O atoms, which in turn gives a ratio of organic mass to organic carbon mass (OM/OC). The OM/OC then allows us to predict the type of compounds contained within the LV-OOA. The analysis for Hedo indicated that the selected ions accounting for ~70% of the LV-OOA mass had a C:H:O ratio of 1: 2.6: 1.9, which gave an OM/OC of 3.8 µg µgC$^{-1}$. Similarly, the C:H:O ratios determined for the LV-OOA at Fukue and Fukuoka were 1: 2.6: 1.8 and 1: 2.6: 1.7, respectively, giving OM/OC values of ~ 3.6 µg µgC$^{-1}$ and ~ 3.5 µg µgC$^{-1}$, respectively. These C:H:O ratios were very similar and gave OM/OC values similar to fulvic and humic acids (known as humic-like substances or HULIS) analyzed in the laboratory.[26] The LV-OOA present at the three sampling sites was likely composed of HULIS. The C:H:O ratios for the HOA at Fukue and Fukuoka were 1:2.1:0 and 1:1.9:0, which both gave an OM/OC of 1.2 µg µgC$^{-1}$. Using the quantitative results for LV-OOA and HOA provided in Table S-1, the overall OM/OC ratios at Fukue and Fukuoka were estimated to be 2.8 µg µgC$^{-1}$ and 2.5 µg µgC$^{-1}$, respectively.

**Comparison of LV-WSOC and OA concentrations.** A comparison of 24-h average LV-WSOC concentrations with 24-h average *m/z* 44 mass concentrations showed a clear positive relationship (Figure 2), indicating an association between the carboxylic acids in fine OA and the LV-WSOC in TSP. Linear regression demonstrated that the slopes at Hedo and Fukue were equivalent within the range of uncertainties, whereas the slope at Fukuoka was twice as large than the other two sites. This was possibly because at Fukue and Hedo, the LV-WSOC consisted of HULIS as indicated by the LV-OOA factor of the PMF analysis, whereas at Fukuoka, the LV-WSOC was composed of HULIS and/or semi-volatile carboxylic acids. Assuming that the proportional increases at Hedo and Fukue were predominantly due to the contribution of HULIS, the division of the slope in Figure 2 by the slope in Figure S-5 gives a concentration ratio of fine OA to LV-WSOC (OA/LV-WSOC), which is comparable with the OM/OC ratios. The comparison produced OA/LV-WSOC ratios of 0.87 and 1.5 at Hedo and Fukue, respectively. These values are approximately a quarter and a half of the overall OM/OC ratios at Hedo and Fukue, respectively. This discrepancy likely indicates that a significant amount of LV-WSOC existed in the larger particle size than in the $PM_{1.0}$ analyzed by the AMSs. However, the strong correlations between the LV-WSOC and the m/z 44 suggest a shift in the SOA size distribution, possibly due to coagulation.

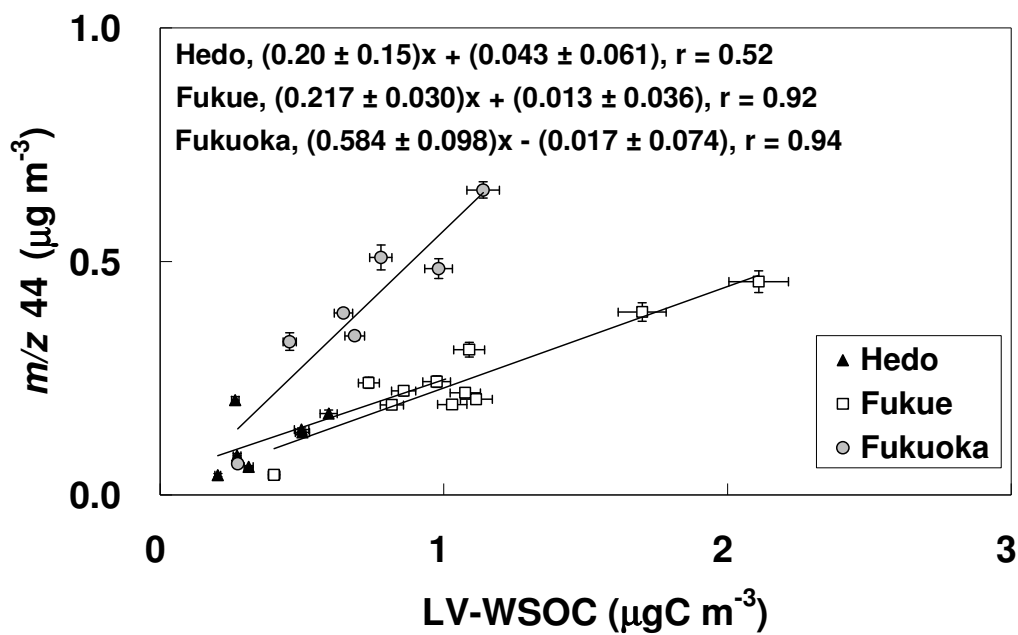

*Figure 2.* Scatter plot of 24-h average m/z 44 concentration versus 24-h average LV-WSOC concentration. The equations shown in the figure are for the linear regressions presented in the figure.

**δ¹³C, $f_{44}$, and $t$[OH].** The high precision $\delta^{13}C$ measurements captured unique features of LV-WSOC: plots of $\delta^{13}C$ of LV-WSOC versus $f_{44}$ clearly indicate an increasing systematic change as $f_{44}$ increases at Hedo, a decreasing systematic change at Fukue, and random variation at Fukuoka (Figure 3). Linear regressions drawn by the least-square method were $\delta^{13}C$ = (24‰ ± 5‰) × $f_{44}$ -(29‰ ± 1‰) with r² = 0.809 for the Hedo data; (-81‰ ± 34‰) × $f_{44}$ -(12‰ ± 4‰) with r² = 0.392 for the Fukue data; and (14‰ ± 40‰) × $f_{44}$ -(24‰ ± 3‰) with r² = 0.025 for the Fukuoka data. Note that the uncertainties shown are standard errors. These results indicate that $\delta^{13}C$ and $f_{44}$ are strongly correlated at Hedo with a significantly positive slope, less correlated with a significantly negative slope at Fukue, and there is almost no correlation with an insignificant slope at Fukuoka. The $\delta^{13}C$ and the $f_{44}$ values were further evaluated by comparison with another photooxidation indicator, $t$[OH].

Assuming that the irreversible reaction of $NO_2$ with the OH radical, which produces $HNO_3$, is the major conversion pathway of $NO_x$ to $NO_y$, $t$[OH] (a product of the average reaction time of $NO_x$, $t$, × the average concentration of OH radical, [OH]) can be described as follows using its second-order rate law[28,29]:

$$t[\text{OH}] = -\frac{1}{k_{NO_2}} \ln \frac{[\text{NO}_x]}{[\text{NO}_y]} \qquad (1)$$

where $[NO_x]$, $[NO_y]$, and $k_{NO2}$ are the concentrations of $NO_x$ and $NO_y$ in molecules cm⁻³ at time $t$ and the rate constant for the reaction of $NO_x$ with OH radical, respectively.

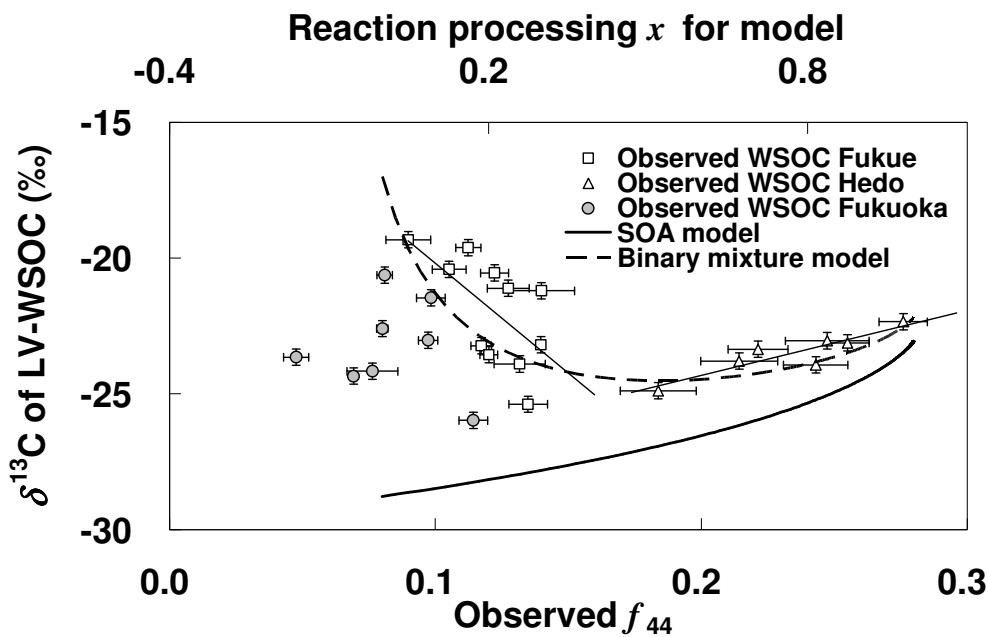

*Figure 3.* Scatter plot of the observed $\delta^{13}C$ of LV-WSOC versus the observed 24-h average $f_{44}$ of OA (lower horizontal axis), and a modeled plot of $\delta^{13}C$ for SOA only (solid curve) and for a binary mixture of SOA with background LV-WSOC (dashed curve) as a function of precursor reaction processing, x (upper horizontal axis). The straight lines are linear regressions for the observed data points at Fukue and Hedo. See the text for details of the model calculation and the linear regressions.

We used a $k_{NO2}$ value of $8.7 \times 10^{-12}$ cm$^3$ molecule$^{-1}$ s$^{-1}$ at 300 K and 1 atm[30, (and references therein)] to determine $t$[OH]. The estimated $t$[OH] ranged from $2.0 \times 10^7$ to $4.0 \times 10^7$ h$^{-1}$ molecule$^{-1}$ cm$^{-3}$. With $5 \times 10^5$ h$^{-1}$ molecule$^{-1}$ cm$^{-3}$ as used by Takegawa et al.[28], we estimated that the reaction time (*t*) would be 40 to 80 h, which was within the range of the transport times roughly estimated from the back trajectory analysis discussed earlier. A plot of $f_{44}$ as a function of $t$[OH] showed a slightly curving form (Figure 4), which was the typical shape of the $f_{44}$ plot observed for SOA.[31] Its linear approximation gave a slope of $(2.1 \pm 0.5) \times 10^{-9}$ h$^{-1}$ molecule$^{-1}$ cm$^3$ (r$^2$ = 0.635). This value in turn gives the average rate of increase of $f_{44}$ as $1.1 \times 10^{-3}$ per hour, if the [OH] value referred to earlier is used. A plot of $\delta^{13}$C against $t$[OH] showed a weak correlation (r$^2$ = 0.141), but with a consistently negative slope of -1.3 ± 1.0 ‰ h$^{-1}$ molecule$^{-1}$ cm$^3$ with an intercept of -18.6 ± 2.8 ‰ (Figure 4). This decreasing trend was consistent with the trend observed at Fukue, as shown in Figure 3. The qualitative agreement between the $\delta^{13}$C profiles plotted against two different oxidation indicators suggested their validity. As discussed later for simple $\delta^{13}$C models, the $\delta^{13}$C values for the background LV-WSOC were also very similar. We then considered the following possibilities for the two different systematic variations: $\delta^{13}$C of LV-WSOC would systematically change when, (1) two primary LV-WSOCs that have a different combination of fixed $\delta^{13}$C and $f_{44}$ values are mixed (i.e., two-endpoint mixing), (2) isotope fractionations occur during the condensation or evaporation of low-volatile compounds, (3) isotope fractionations occur during the oxidation reaction inside particles, or (4) isotope fractionations occur during the oxidation reaction of SOA precursor gases.

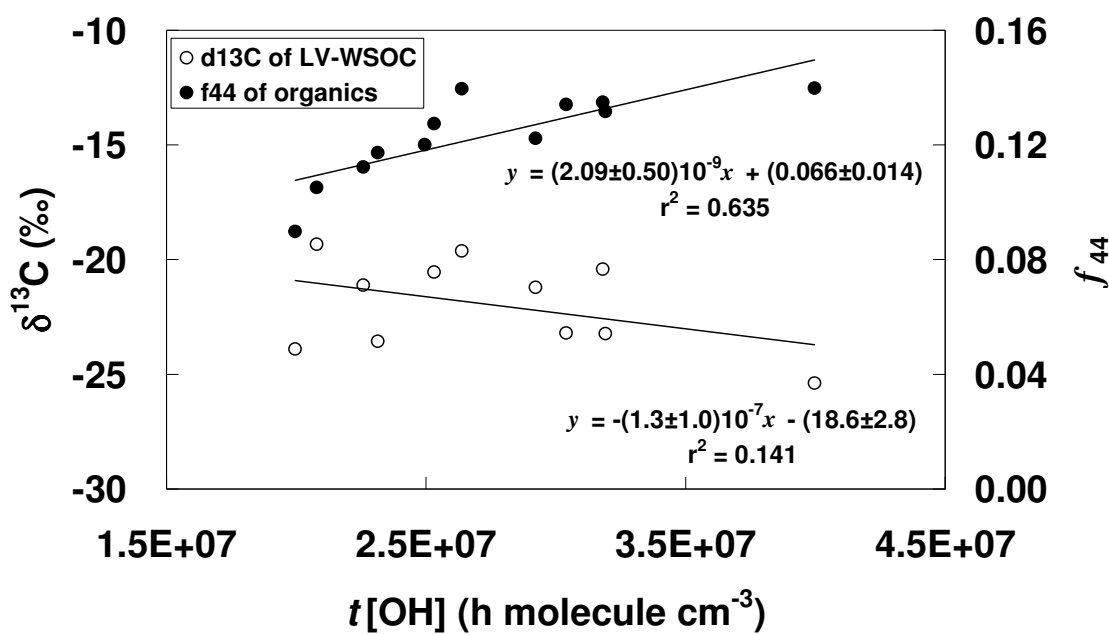

***Figure 4.*** *Scatter plot of $\delta^{13}C$ of LV-WSOC (left axis) and the 24-h average $f_{44}$ of OA (right axis) at Fukue as a function of photochemical age (t[OH]) estimated by the $NO_x/NO_y$ ratio. The straight lines are linear regressions for the $\delta^{13}C$ and the $f_{44}$ plots.*

Two-endpoint mixing changes the $\delta^{13}C$ of an LV-WSOC mixture systematically as a fraction of one of the two members gradually increases. If the observations at Hedo were modeled as the mixture of a minor fraction of "type A" LV-WSOC with low $f_{44}$ and low $\delta^{13}C$ (e.g., $f_{44} = 0.06$ and $\delta^{13}C = -30‰$) and a major fraction of "type B" LV-WSOC with high $f_{44}$ and high $\delta^{13}C$ (e.g., $f_{44} = 0.3$ and $\delta^{13}C = -17‰$), then $\delta^{13}C$ of the LV-WSOC mixture increases with the fraction of type B. In contrast, to reproduce the observations at Fukue would require a mixture where a minor fraction of "type C" LV-WSOC with low $f_{44}$ and high $\delta^{13}C$ (e.g., $f_{44} = 0.06$ and $\delta^{13}C = -17‰$) mixes with a major fraction of "type D" LV-WSOC with high $f_{44}$ and low $\delta^{13}C$ (e.g., $f_{44}$ = 0.3 and $\delta^{13}C$ = –30‰). $\delta^{13}C$ of the mixture decreases as the fraction of type D LV-WSOC increases. Therefore, to explain the systematic variations using the two-endpoint mixing model, consistent contributions of the four types of LV-WSOC are needed. To our knowledge, combinations of $f_{44}$ and $\delta^{13}C$ for type A and B are possible if they originate from $^{13}C$ depleted primary organics[10] and SOA that has been completely converted from VOCs of $C_4$ plant origin, respectively. Combinations of $f_{44}$ and $\delta^{13}C$ for type C and D LV-WSOC are feasible as primary organics from $C_4$ plant biomass burning[10, 32, 33] and SOA from the photooxidation of anthropogenic VOC mixtures[11,12]. Considering the significant influence of transboundary anthropogenic pollutants, as indicated by the presence of sulfate, as well as the fact that biogenic VOCs rapidly react with both ozone and OH radical, it is not feasible that there was a more significant impact from the on-going production of SOA from biogenic VOCs at Hedo than at Fukue.

Evaporative or condensate carbon and hydrogen isotope fractionation of organic substances [34–46] also systematically changes $\delta^{13}C$ of condensed organics. Depending on the substance, the direction of fractionation is toward heavy or light isotopes. However, regardless of the direction of fractionation the degree of fractionation reported is small. For example, Irei[37] observed only +0.3‰ change in $\delta^{13}C$ after evaporating ~11% of a laboratory-derived SOA carbon sample with a 6.3 L min$^{-1}$ flow of dry air for 24 h at room temperature. Given that the WSOC we studied had "low-volatility," it is unlikely that this mechanism could account for the variations in $\delta^{13}C$ that we observed.

Oxidation in the condensed phase, referred to as particle aging, may be an explanation for the changes in $f_{44}$ and $\delta^{13}C$. It has been observed in laboratory studies of SOA that the $f_{44}$ of SOA increases without the production of SOA, which is indicative of the production of carboxylic acids inside the particles.[3] This oxidation process would have resulted in forward isotope fractionations. However, it has been reported that oxalic acids are photocatalytically oxidized in the condensed phase (i.e., oxidative loss) and the reaction causes forward carbon isotope fractionation.[38] These reactions are counter-processing for carboxylic acids, and the reports suggest that carboxylic acids are not only oxidation products, but may also act as reactants. It should be noted that oxidation in the condensed phase is considerably slower than that in the gas-phase.[39,40] Assuming that the two types of particle aging are the explanations for the opposite systematic $\delta^{13}C$ changes observed, the former process must be occurring at Hedo and the latter must be occurring at Fukue. However, this is inconsistent with our expectation that the latter process would be more important at Hedo, where the transport time was longer. According to the back trajectories of air masses (Figure S-2), the transport time from the continent to Hedo was 6 to 80 h longer than to Fukue. This significant difference in transport time can also be seen in the concentrations of components. Nevertheless, no indication of oxidative loss was found in $\delta^{13}C$ at Hedo, and therefore there was probably also no loss in the samples at Fukue.

Isotopic fractionation occurring during the oxidation reaction of a precursor gas (i.e., a kinetic isotope effect (KIE) when SOA is formed is a plausible explanation for our observations. A stable carbon KIE for compounds containing $^{13}$C at natural abundance levels is defined as the ratio of the reaction rate constants for a reactant containing only $^{12}$C atoms ($k_{12}$) and the same reactant containing $^{13}$C atoms ($k_{13}$), and in this study it is expressed as $\varepsilon = (k_{12}/k_{13}) - 1$. Irei et al.[11,12] reported that the difference in $\delta^{13}$C between precursor toluene and SOA formed by the photooxidation of toluene (the precursor $\delta^{13}$C minus the SOA $\delta^{13}$C) ranged from –3‰ to –6‰, varying systematically with the extent of the oxidation reaction. This isotope fractionation was caused by KIEs that occurred during the series of oxidation reactions leading to SOA formation. The resulting $\delta^{13}$C profile resembles the systematic $\delta^{13}$C variation observed at Hedo (Figure 3). This explanation is also consistent with the fact that the OA and the LV-WSOC concentrations increased with the m/z 44 concentrations at all sites (Figure S-4 and Figure 2). The consistent results obtained in this study suggest that $f_{44}$ can be used as an indicator of the extent of the precursor reaction. However, its use would probably be limited to cases where there are reaction products with a unique $f_{44}$ mix, such as a constant amount of organics each with significantly different $f_{44}$ values (i.e., a binary mixture with different values of $f_{44}$, such as a mixture of LV-OOA and HOA). There is another possible limitation in the use of $f_{44}$ as an indicator of a reaction process, when the $f_{44}$ becomes saturated by SOA predominantly composed of a single substance, which has a constant $f_{44}$.

This saturation hypothesis seems to contradict the results of the PMF analysis discussed earlier where ~ 99% of the OA at Hedo was composed of LV-OOA. It is possible that the PMF analysis could not deconvolute the small contribution of the HOA factor or the systematic $\delta^{13}$C variation

We further compared the observed $\delta^{13}$C changes with modeled $\delta^{13}$C variations using a simple calculation for a binary mixture of SOA, which was subjected to irreversible isotope fractionation and a constant amount of background LV-WSOC, with a distinctive $f_{44}$ and $\delta^{13}$C. The change in $\delta^{13}$C of SOA ($\delta^{13}C_{SOA}$) can be modeled given the KIE for the reaction that forms SOA from a precursor, the initial $\delta^{13}$C of the precursor, and the extent of precursor reaction processing. We modeled a $\delta^{13}C_{SOA}$ profile (Figure 3) according to a calculation presented by Irei et al.,[12] using values for the epsilon expression ($\varepsilon$) of the KIE of 6‰ and the initial $\delta^{13}$C for a precursor ($^0\delta^{13}C_p$) of –23‰, which we adopted from the carbon KIE for the reaction of toluene with the OH radical[41] and the initial $\delta^{13}$C of carbonaceous aerosols from the combustion of fossil fuels used in Asia,[33,42] respectively. The 6‰ $\varepsilon$ is within the range of typical KIEs (2‰ to 19‰) that occur during the atmospheric oxidation of the major airborne VOCs at ground level.[6] We also modeled a $\delta^{13}$C profile for a binary mixture of SOA accumulating against a constant amount of background LV-WSOC ($\delta^{13}C_{binary}$), in which the $\delta^{13}$C of the background LV-WSOC ($\delta^{13}C_{bkg}$) was -17‰. The calculation used to determine $\delta^{13}C_{binary}$ was based on the following mass balance:

$$\delta^{13}C_{binary} = (1-w_{SOA}) \times \delta^{13}C_{bkg} + w_{SOA} \times \delta^{13}C_{SOA}, \quad (2)$$

where $w_{SOA}$ is the mass fraction of SOA in the binary mixture, which is a dependent variable of the extent of precursor reaction processing $x$. Note that the sum of the mass fractions of the background LV-WSOC and SOA must be unity. Equation (2) was then combined with the Rayleigh-type function used to determine the $\delta^{13}C_{SOA}$ as follows:

$$\delta^{13}C_{binary} = \left\{ (1-w_{SOA}) \times (\delta^{13}C_{bkg} + 1) + w_{SOA} \times \frac{{}^0\delta^{13}C_p + 1}{x} \times \left[1 - (1-x)^{\left(\frac{1}{1+\varepsilon}\right)}\right] \right\} - 1. \quad (3)$$

The dependent variable $w_{SOA}$ in (3) can be expressed as a function of SOA yield ($Y_{SOA}$), which is another variable proportional to the independent variable $x$.[12] Assuming that the conditions of the laboratory studies are applicable to the atmosphere, $w_{SOA}$ can be expressed as:

$$w_{SOA} = \frac{Y_{SOA} \cdot b}{a + Y_{SOA} \cdot b} = \frac{(0.3x - 0.025) \cdot b}{a + (0.3x - 0.025) \cdot b}, \quad (4)$$

where $a$ and $b$ are arbitrary constants ($\mu gC\ m^{-3}$) determining the carbon mass concentrations for background LV-WSOC and the SOA yielded in the binary mixture, respectively. Note that the right side of eq (4) must be less than or equal to unity. For the model calculation, we used 0.05 $\mu gC\ m^{-3}$ and 1 $\mu gC\ m^{-3}$ for $a$ and $b$, respectively, and regarded $0.025b$ as a negligibly small value. By combining eqs (3) and (4), we yielded an equation with a variable $x$ and five parameters ($\varepsilon$, $\delta^{13}C_{bkg}$, ${}^0\delta^{13}C_p$, $a$, and $b$). Using the parameter values given earlier, $\delta^{13}C_{binary}$ was calculated (Figure 3). As a comparison, $\delta^{13}C_{binary}$ calculated using different parameter values are shown in Figure S-10.

A qualitative comparison between the modeled and observed $\delta^{13}$C profiles revealed that the trend for increasing $\delta^{13}$C observed at Hedo was consistent with both the modeled $\delta^{13}$C$_{SOA}$ and $\delta^{13}$C$_{binary}$ profiles, which had undergone extensive oxidation reaction processing. This suggests that SOA was more likely to be the major component of LV-WSOC observed at Hedo, and the unknown precursor(s) contained in the air masses from continental China likely had a $\delta^{13}$C$_{bkg}$ value between -17‰ and -13‰ (Figure S-10a), an ε value between 3‰ and 6‰ (Figure S-10c), and a $^0\delta^{13}$C$_p$ value between –23‰ and -18‰ (Figure S-10d). The binary SOA mixture model also successfully reproduced the trend for a decreasing $\delta^{13}$C that was observed at Fukue. To be consistent, we applied the binary mixture model to the $f_{44}$ values (see the calculation presented in the Supporting Information), and found that it could also reproduce the $f_{44}$ profile plotted against $t$[OH] (Figure S-11). The series of consistent results led us to conclude that the cumulative contribution of SOA to the background LV-WSOC became more significant in TSP and fine OA as the precursor reaction proceeded. Additionally the binary mixture model suggested that $\delta^{13}$C$_{bkg}$ was -17‰ or higher and that the $f_{44}$ of LV-WSOC was 0.06 or lower. A similar value of $\delta^{13}$C for background total carbon in particulate matter (the intercept of the fit) was observed at Mt. Tai, China.[43] Marine atmospheric studies in the Indian and Pacific Oceans have also documented a significant effect of the emissions from $^{13}$C-enriched sources on particulate WSOC[44,45] and on VOC.[46] These observations in the East-Asian region support the existence of the background

LV-WSOC.

The $\delta^{13}C_{bkg}$ and the $f_{44}$ for the background LV-WSOC supply are evidence of a primary emission source associated with the combustion of $C_4$ plants, unless there is an unreported primary emission source within the range of these $\delta^{13}C$ and $f_{44}$ values. To reproduce the magnitude of the observed $\delta^{13}C$ variation using the $\delta^{13}C_{binary}$ model, $\delta^{13}C_{bkg}$ had to be higher than -17‰ and $f_{44}$ had to be approximately 0.06. A combination of such high $\delta^{13}C_{bkg}$ and low $f_{44}$ values (less oxidation processing) is unique and these values suggest that the source of the background LV-WSOC is the primary organics in soil and street dust, combustion related to $C_4$ plants, or water-insoluble organic carbon (WIOC) from marine aerosol. It has been reported that $\delta^{13}C$ values of black carbon from $C_4$ plant combustion and from soil and street dust range between –12‰ and –19‰.[33,42] Given that high-temperature combustion causes only small isotope fractionations,[46,47] the $\delta^{13}C$ values of black carbon are equivalent to the $\delta^{13}C$ values of LV-WSOC from $C_4$ plant combustion, soils, and street dust. $\delta^{13}C$ values of –20‰ to –22‰ have also been reported in WIOC from marine aerosols.[48] If a portion of this WIOC is dissolved or converted (oxidized) to WSOC without isotope fractionation, then WIOC can also be a possible source of background LV-WSOC.

Most of these candidate sources can be rejected. Local soil and street dust consists of coarse particles (~10 μm) with short atmospheric lifetimes[28] and their concentration strongly depends on meteorological conditions as well as traffic and other human activities.

Considering the large variation in observed hourly wind speeds (80th percentile hourly wind speeds at Fukue and Hedo were 4 m s$^{-1}$), the constancy in the amounts of background LV-WSOC during the study period, and the typically very low levels of traffic near the rural sites, soil and street dust cannot be the source of the background LV-WSOC. If WIOC were the origin, the portion of it converted to LV-WSOC would need to be enriched in $^{13}$C (i.e., significantly large inverse isotope fractionations at the dissolution or oxidation of WIOC) to raise the $\delta^{13}$C of the background LV-WSOC above -17‰. However, such inverse carbon isotope fractionation has not been reported and seems implausible. In contrast, we found no evidence to rule out emissions from C$_4$ plant sources. The $f_{44}$ values as low as 0.06, an indication of primary emission, are consistent with the observed $f_{44}$ of OA from biomass burning (0.16 or less).[49] Our findings and other studies in East Asia[43,50,51] are compatible with primary emissions related to biomass burning. Two species of LV-WSOC, each of which has significantly different $\delta^{13}$C and $f_{44}$, were found mixed together. As demonstrated by Song et al.,[52] it is difficult to separate two species if either of their $\delta^{13}$C or $f_{44}$ values are similar. Long-term observations are needed to evaluate the relationship between $f_{44}$ and oxidation reaction processing, SOA formation, as well as source identification and the variability of the background LV-WSOC in the East-Asian atmosphere.

**ACKNOWLEDGMENTS**

We thank Akio Togashi from the National Institute for Environmental Studies and the students from Tokyo University of Agriculture and Technology, Fukuoka University, and the University of the Ryukyus for their help in the collection of filter samples. We also acknowledge the NOAA Air Resources Laboratory (ARL) for the provision of the HYSPLIT transport and dispersion model and/or READY website (http://www.ready.noaa.gov). This project was financially supported by the Environment Research and Technology Development Fund of the Ministry of Environment, Japan (B-1006 and A-1101), a Grant-in-Aid for Scientific Research on Innovative Areas (No. 4003) from the Ministry of Education, Culture, Sports, Science and Technology, Japan, and partially supported by the International Research Hub Project for Climate Change and Coral Reef/Island Dynamics of Univ. of the Ryukyus.

**Supporting Information Available**

This information is available free of charge via the Internet at http://pubs.acs.org/

various types of combustion-related aerosol particles. *Int. J. Mass Spectrom.* **2006**, *258*, 37-49.

(51) Cheng, Y.; Engling, G.; He, K.-B.; Duan, F.-K.; Ma, Y.-L.; Du, Z.-Y.; Liu, J.-M.; Zheng, M.; Weber, R.J. Biomass burning contribution to Beijing aerosol. Atmos. Chem. Phys. **2013**, 13, 7765-7781.

(52) Song, J.; He, L.; Peng, P.; Zhao, J.; Ma, S. Chemical and isotopic composition of humic-like substances (HULIS) in ambient aerosols in Guangzhou, South China. *Aerosol Sci. Technol.* **2012**, 46, 533-546.

Supporting information to the article entitled

# Transboundary secondary organic aerosol in western Japan indicated by the $\delta^{13}$C of water-soluble organic carbon and the *m/z* 44 signal in organic aerosol mass spectra

Satoshi Irei,[*,1] Akinori Takami,[1] Masahiko Hayashi,[2] Yasuhiro Sadanaga,[3] Keiichiro Hara,[2] Naoki Kaneyasu,[4] Kei Sato,[1] Takemitsu Arakaki,[5] Shiro Hatakeyama,[6] Hiroshi Bandow,[3] Toshihide Hikida,[7] and Akio Shimono[7]

[1]National Institute for Environmental Studies, 16-2 Onogawa, Tsukuba, Ibaraki 305-8506, Japan
[2]Department of Earth System Science, Faculty of Science, Fukuoka University, 8-19-1 Nanakuma, Jonan-ku, Fukuoka 814-0180, Japan
[3]Department of Applied Chemistry, Graduate School of Engineering, Osaka Prefecture University, 1-1 Gakuencho, Naka-ku, Sakai, Osaka 599-8531, Japan
[4]National Institute of Advanced Industrial Science and Technology, 16-1 Onogawa, Tsukuba, Ibaraki 305-8569, Japan
[5]Department of Chemistry, Biology and Marine Science, Faculty of Science, University of the Ryukyus, 1 Senbaru, Nishihara, Okinawa 903-0213, Japan
[6]Agricultural Department, Tokyo University of Agriculture and Technology, 3-5-8 Saiwai-cho, Fuchu, Tokyo 183-8509, Japan
[7]Shoreline Science Research Inc., 3-12-7 Owada-machi, Hachioji, Tokyo 192-0045, Japan

*Corresponding author: Satoshi Irei, National Institute for Environmental Studies, 16-2 Onogawa, Tsukuba, Ibaraki 305-8506, Japan (phone: +81-29-850-2314; fax: +81-29-850-2579; e-mail: satoshi.irei@gmail.com)

**9 Pages**
**1 Table**
**11 Figures**

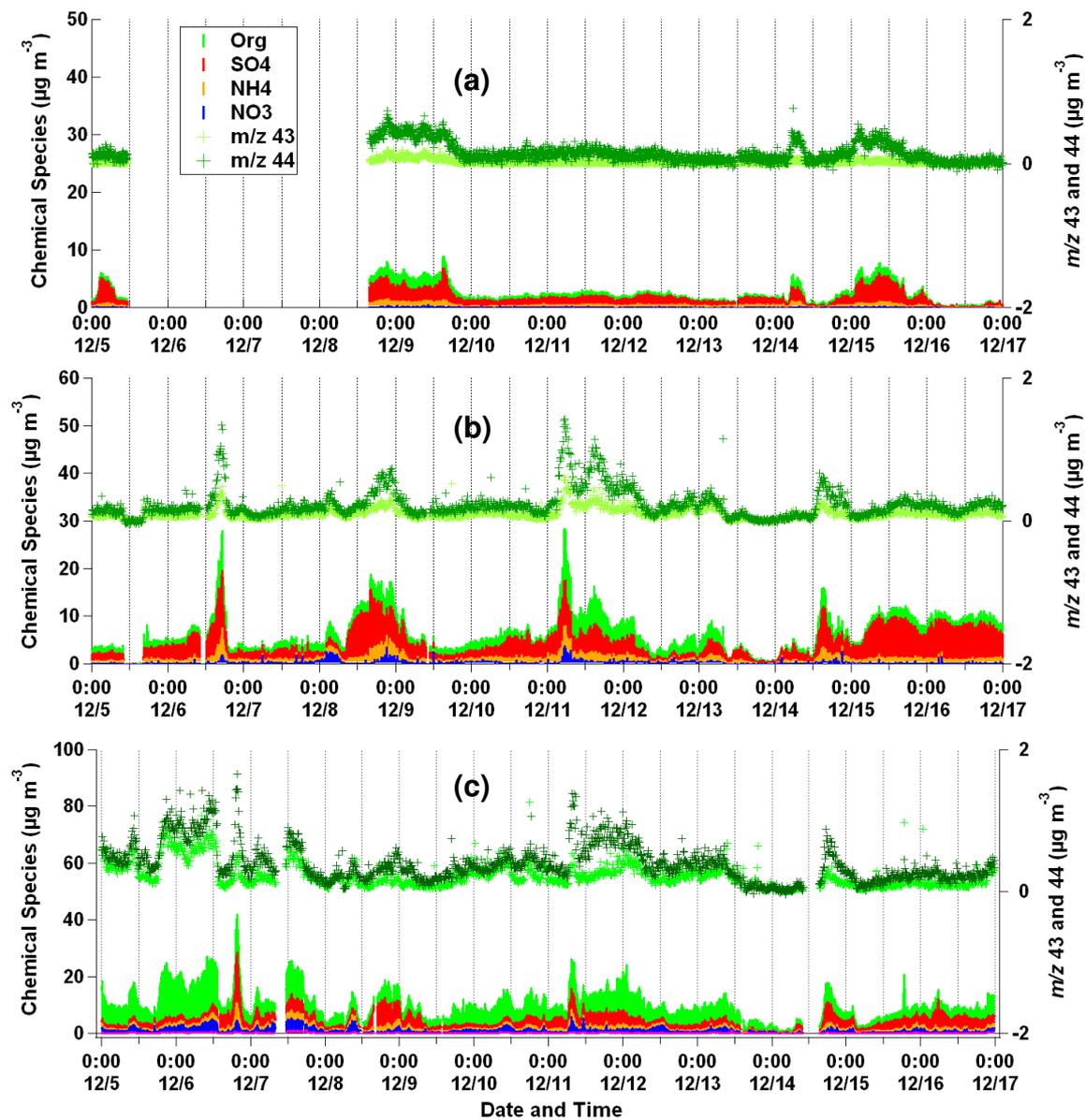

**Figure S-1.** Time series plot of chemical species concentrations (left axis) and m/z 43 and 44 concentrations (right axis) measured by ACSM and AMS at (a) Hedo, (b) Fukue, and (c) Fukuoka.

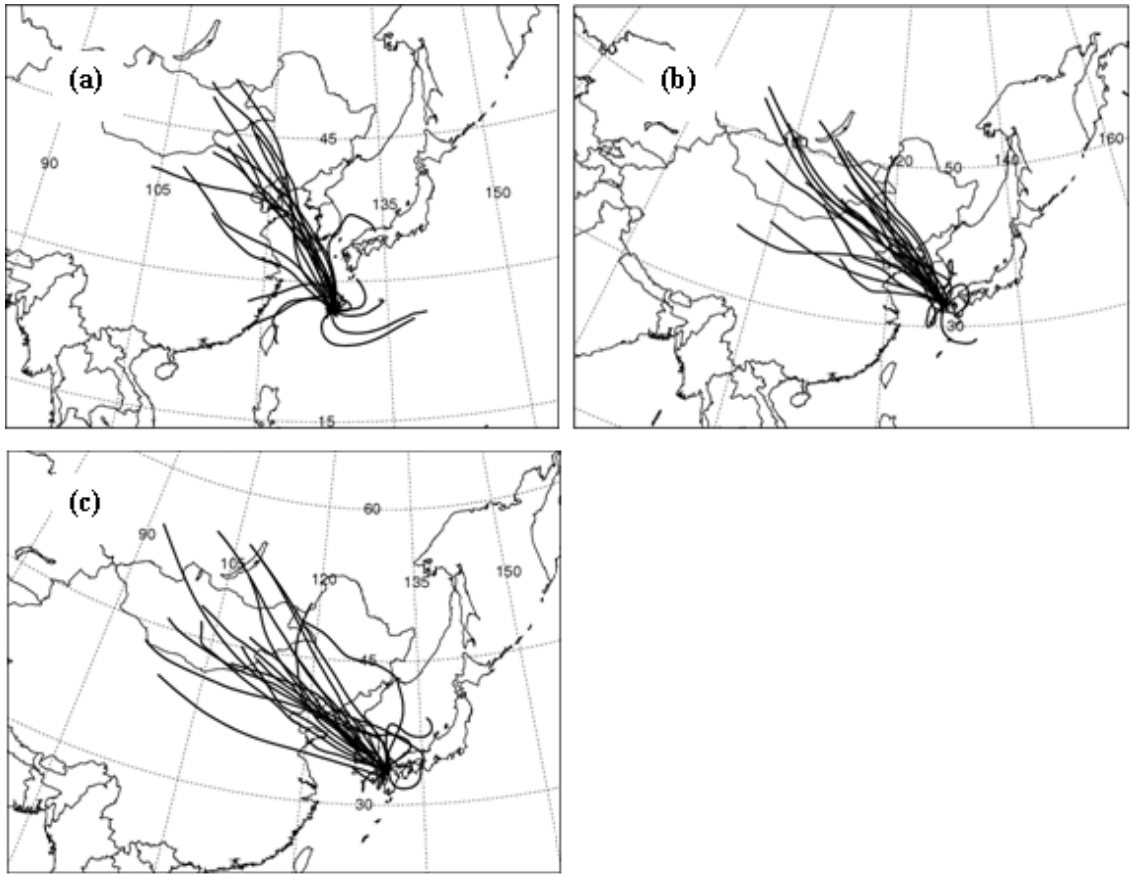

**Figure S-2.** 48 hour back trajectories of 500 m AGL air masses arriving at (a) Hedo, (b) Fukue, and (c) Fukuoka during the study period (December 6th to 17th). The trajectories were drawn every 12 hour.

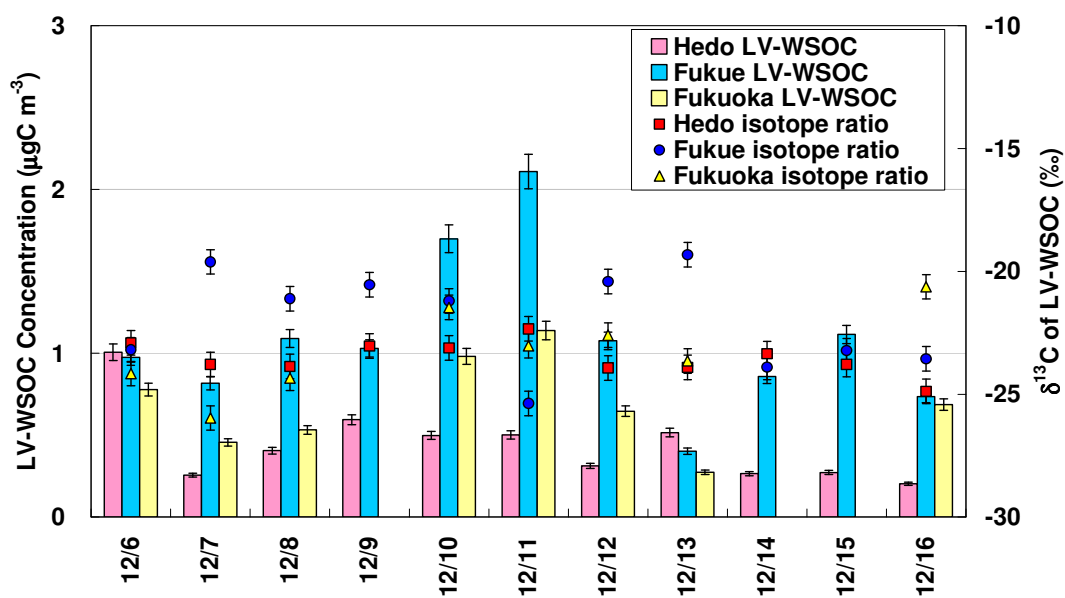

**Figure S-3.** Time series plot of concentration (left axis) and $\delta^{13}C$ (right axis) of LV-WSOC observed at Hedo, Fukue, and Fukuoka.

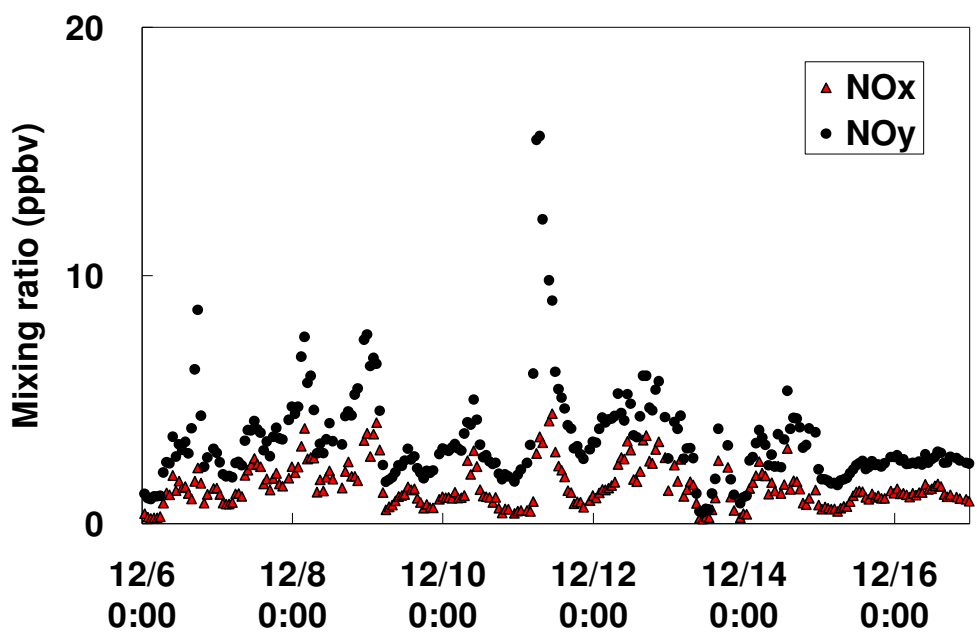

**Figure S-4.** Time series plot of hourly average mixing ratio for $NO_x$ and $NO_y$ at Fukue.

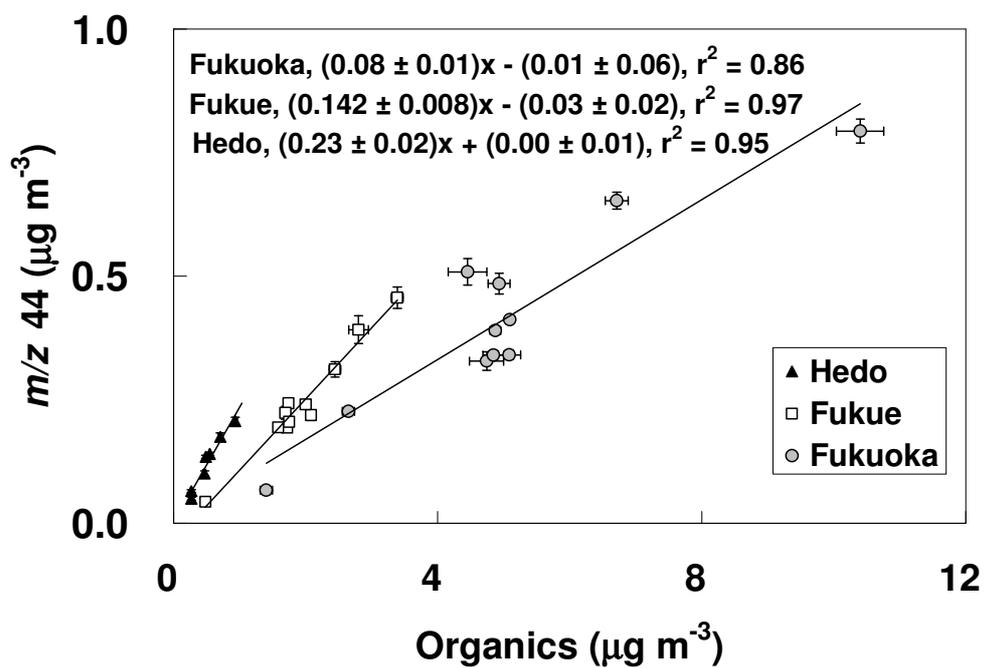

**Figure S-5.** Scatter plot of 24-h average m/z 44 concentration versus 24-h average organic aerosol concentration.

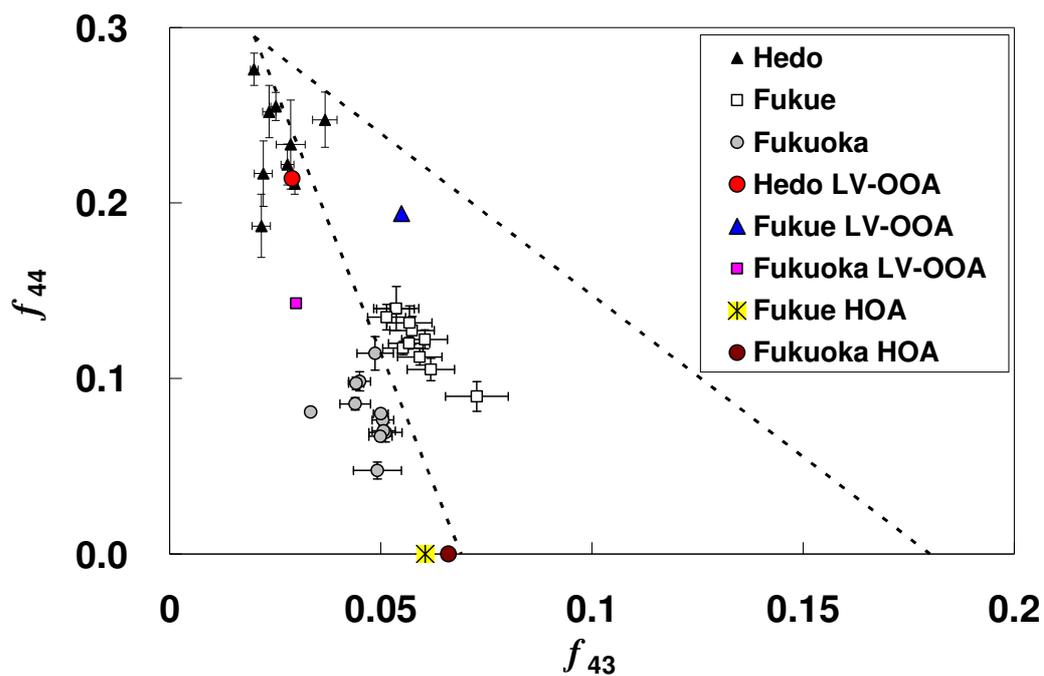

**Figure S-6.** Scatter plot of 24-h average $f_{44}$ of OA versus 24-h average $f_{43}$ of OA. Dashed lines are the limits of oxidation states reported by Ng et al[2] (see the text for the detail of this discussion). $f_{44}$ and $f_{43}$ for LV-OOA and HOA factors yielded by PMF analysis are also shown.

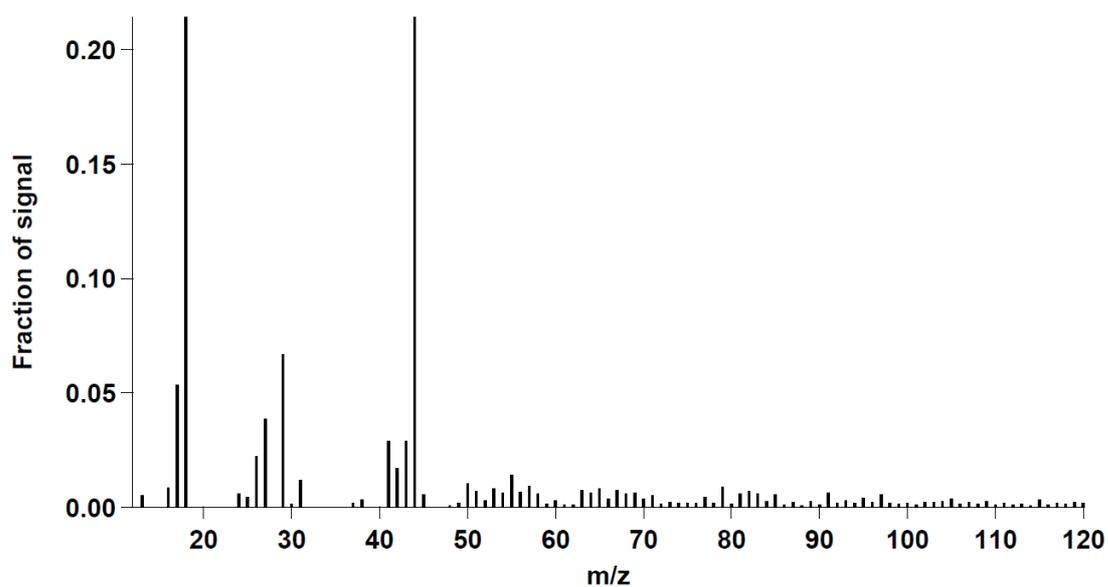

**Figure S-7.** Loading (m/z 10 to m/z 120) yielded from one factorial PMF analysis on organic mass spectra obtained at Hedo. The loading was identified as LV-OOA.

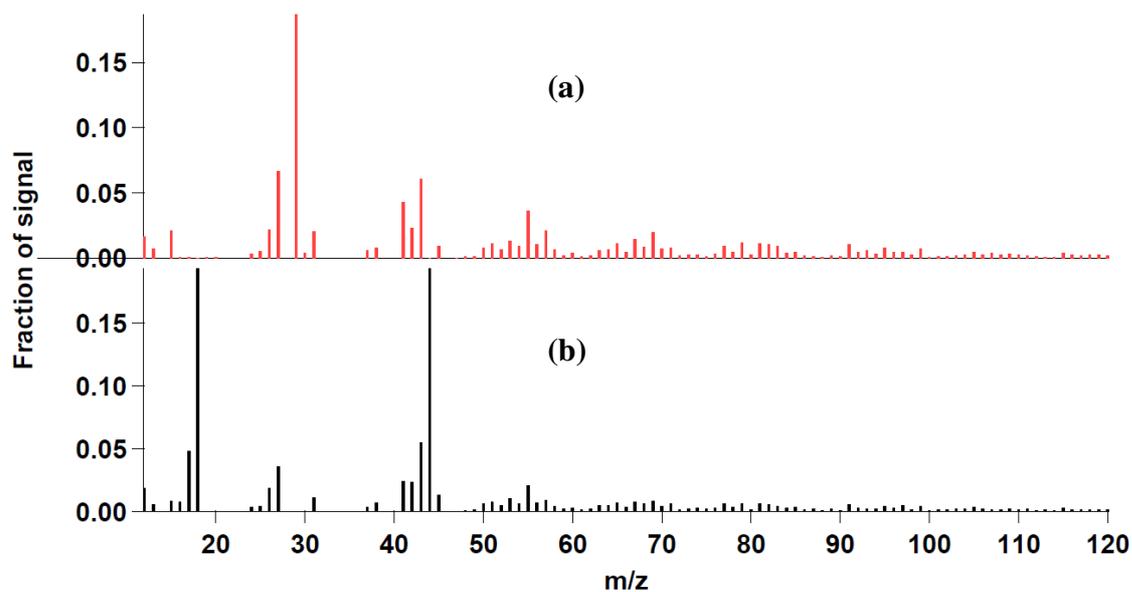

**Figure S-8.** Loading (m/z 10 to m/z 120) yielded from two factorial PMF analysis on organic mass spectra obtained at Fukue. The loadings were identified as (a) HOA and (b) LV-OOA.

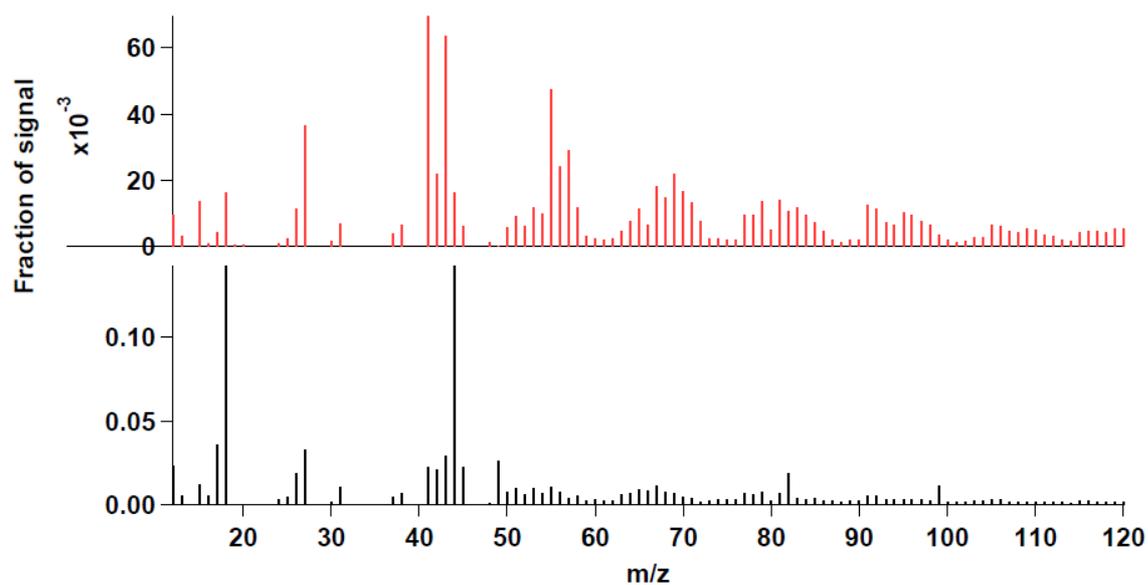

**Figure S-9.** Loading (m/z 10 to m/z 120) yielded from two factorial PMF analysis on organic mass spectra obtained at Fukuoka. The loadings were identified as (a) HOA and (b) LV-OOA.

**Table S-1.** Summary for the fractions of yielded factors by PMF analysis

| Site | Type of OA | Mass Fraction (%) |
|---|---|---|
| Hedo[a] | LV-OOA | 99 |
| Fukue[b] | HOA | 32 |
|  | LV-OOA | 67 |
| Fukuoka[b] | HOA | 44 |
|  | LV-OOA | 56 |

[a]Results of one factorial analysis. [b]Results of two factorial analysis.

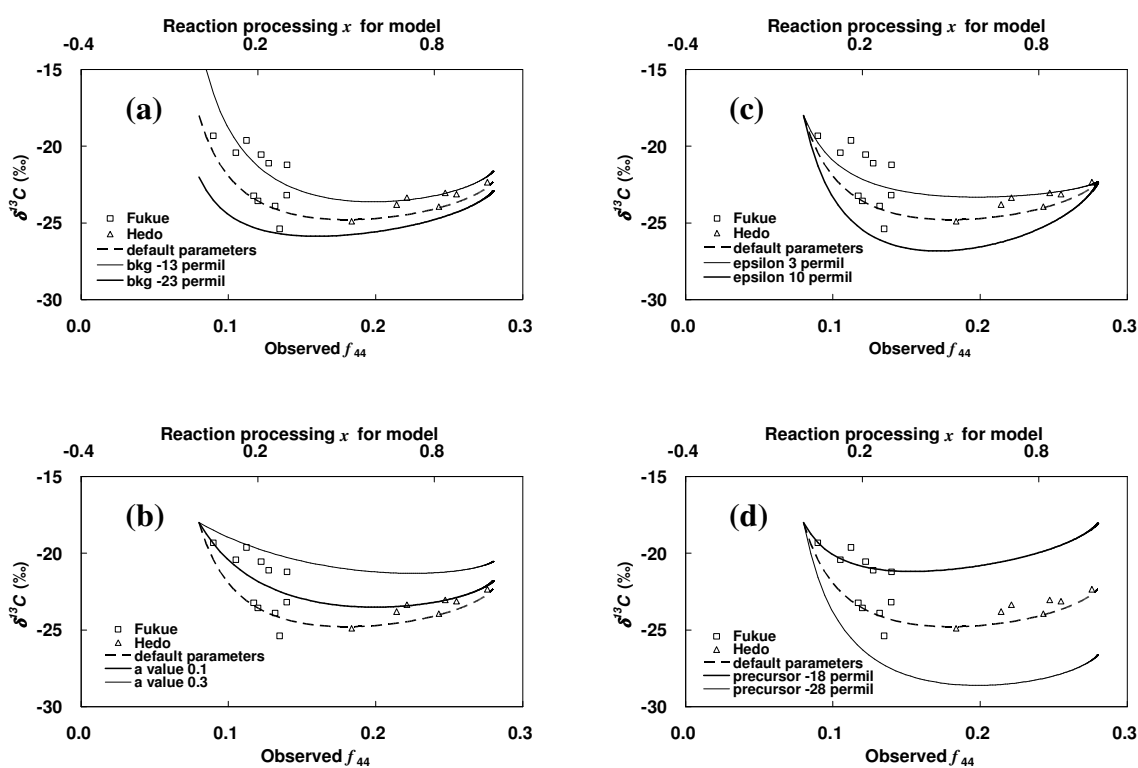

**Figure S-10.** Plot of modeled $\delta^{13}C$ for a binary mixture of SOA ($\delta^{13}C_{binary}$) with background LV-WSOC using (a) different $\delta^{13}C_{bkg}$ values, (b) different "$a$" values, (c) different $\varepsilon$ values, and (d) different $^0\delta^{13}C_p$ values. "Default parameters" used for $\delta^{13}C_{bkg}$, $^0\delta^{13}C_p$, $a$, and $b$ values are, -18‰, -23‰, 0.05 μgC m$^{-3}$, and 1 μgC m$^{-3}$, respectively. See the text for calculation and explanations for the parameters.

**Binary mixture model for $f_{44}$**

As two different types of OA (e.g., HOA and LV-OOA) are mixed, a fraction of signal at m/z 44 in OA mass spectra ($f_{44}$) can be determined using the following mass balance:

$$f_{44} = \frac{{}^{HOA}f_{44} \cdot m_{HOA} + {}^{LVOOA}f_{44} \cdot m_{LVOOA}}{m_{HOA} + m_{LVOOA}} \qquad \text{s-eq. 1}$$

where ${}^{LVOOA}f_{44}$, ${}^{HOA}f_{44}$, $m_{LVOOA}$, and $m_{HOA}$ are the fraction of m/z 44 signal for the LV-OOA factor, the fraction of m/z 44 signal for the HOA factor, the mass concentration for the LV-OOA, and the mass concentration for the HOA, respectively. This equation can be further extended to the following equation by using a combination of the SOA carbon yield equation referred in the text ($y = 0.3x$, where $x$ is the extent of precursor oxidation reaction), mass to carbon mass ratios for the LV-OOA, (OM/OC)$_{LVOOA}$, and the HOA, (OM/OC)$_{HOA}$ discussed in the text, as well as $a$ and $b$ values used in eq. (4) in the text;

$$f_{44} = \frac{{}^{HOA}f_{44} \cdot a \cdot (\frac{OM}{OC})_{HOA} + {}^{LVOOA}f_{44} \cdot \left[ 0.3x \cdot b \cdot (\frac{OM}{OC})_{LVOOA} \right]}{a \cdot (\frac{OM}{OC})_{HOA} + \left[ 0.3x \cdot b \cdot (\frac{OM}{OC})_{LVOOA} \right]} \qquad \text{s-eq. 2.}$$

Here, we used 0.05 µgC m$^{-3}$, 1 µgC m$^{-3}$, 0.06, 0.22, 1.2, and 3.7 for $a$ value, $b$ value, ${}^{HOA}f_{44}$, ${}^{LVOOA}f_{44}$, (OM/OC)$_{HOA}$, and (OM/OC)$_{LVOOA}$, respectively. Modeled plot for $f_{44}$ of this binary mixture is shown in Figure S-11.

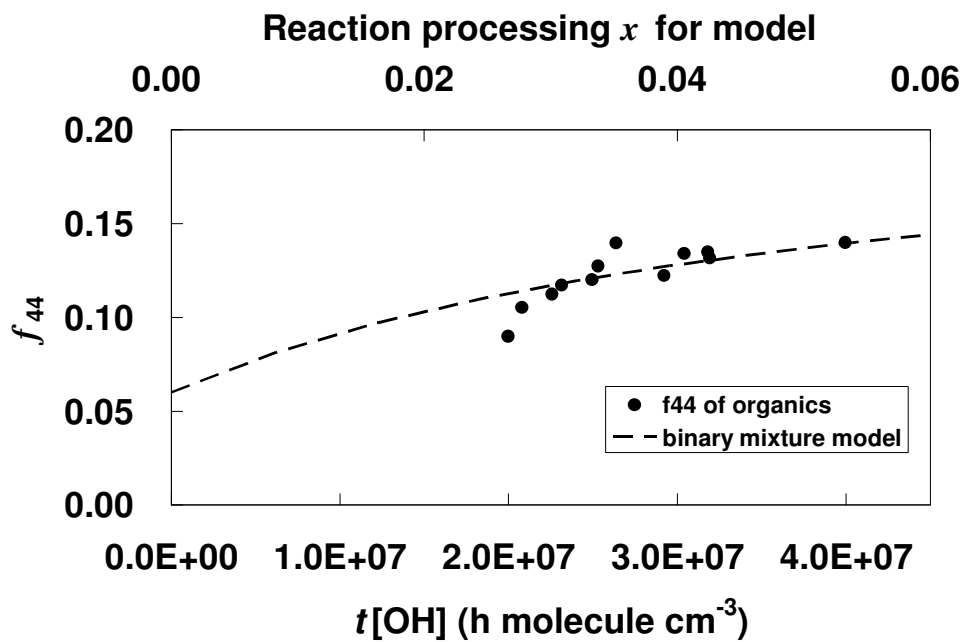

**Figure S-11.** Scatter plot for observed 24-h average $f_{44}$ at Fukue as function of $t$[OH] (solid dark) and modeled plot for $f_{44}$ of binary mixture using s-eq. 2.